\documentclass[twocolumn,aps,showpacs,floatfix,prc]{revtex4}
\usepackage[dvips]{epsfig}
\begin{document}
\title{ Landau quantization and mass-radius relation of magnetized White Dwarfs in general relativity }

\author{ Somnath Mukhopadhyay$^{1*}$, Debasis Atta$^{2*\dagger}$ and D.N. Basu$^{3*}$ }

\affiliation{$^*$ Variable  Energy  Cyclotron  Centre, 1/AF Bidhan Nagar, Kolkata 700 064, India }
\address{ $^\dagger$ Shahid Matangini Hazra Govt. Degree College for Women, Tamluk, West Bengal 721 649, India}

\email[E-mail 2: ]{somnathm@vecc.gov.in}
\email[E-mail 2: ]{debasisa906@gmail.com}
\email[E-mail 3: ]{dnb@vecc.gov.in}
\date{\today }

\begin{abstract}

    Recently, several white dwarfs have been proposed with masses significantly above the Chandrasekhar limit, known as Super-Chandrasekhar White Dwarfs, to account for the overluminous Type Ia supernovae. In the present work, Equation of State of a completely degenerate relativistic electron gas in magnetic field based on Landau quantization of charged particles in a magnetic field is developed. The mass-radius relations for magnetized White Dwarfs are obtained by solving the Tolman-Oppenheimer-Volkoff equations. The effects of the magnetic energy density and pressure contributed by a density-dependent magnetic field are treated properly to find the stability configurations of realistic magnetic White Dwarf stars.

\end{abstract}

\maketitle

\noindent
\section{Introduction}
\label{section1}
    Ultrahigh magnetic fields in nature are known to be associated with compact astrophysical objects namely white dwarfs, neutron stars and black holes. Of these, the largest magnetic fields are found on the surfaces of magnetars, Anomalous X-ray Pulsars (AXPs) and Soft Gamma Repeaters (SGRs), certain classes of neutron stars, with an order of magnitude of $10^{15}$ gauss. Recently, a strong magnetic field of the same order of magnitude as that of a magnetar has been found at the jet base of a supermassive black hole PKS 1830-211 \cite{Vi15}. These strong magnetic fields drastically modify the Equation of State (EoS) of a compact star and its stability. Hence, studying the EoS and equilibria of compact stars in presence of high magnetic fields is an important and rapidly growing field of research in theoretical astrophysics. Magnitudes of magnetic fields of white dwarfs are constrained by the virial theorem:
    
\begin{equation}
\left(\frac{4}{3} \pi R^3\right)\frac{B^2}{8 \pi}=\frac{3}{5}\frac{GM^2}{R},
\label{seqn1}
\end{equation}        
which gives

\begin{equation}
B_{max}=B_{ \odot}\left(\frac{M}{M_{ \odot}}\right)\left(\frac{R}{R_{ \odot}}\right)^{-2}.
\label{seqn2}
\end{equation}
Here, $B,~B_{ \odot},~M,~M_{ \odot},~R,~R_{ \odot}$ are the magnetic field, mass and radius of the white dwarf and sun respectively. Using $B_{ \odot}=2 \times 10^8$ gauss, $M=$ 1.4 $M_{ \odot}$ and $R=$ 0.0086 $R_{ \odot}$, we get the order of magnitude as $B_{max} \sim 10^{12}$ gauss.

    The recently observed peculiar Type Ia supernovae, e.g. SN2006gz, SN2007if, SN2009dc, SN2003fg, \cite{Ho06,Sc10,Hi07,Ya09,Si11} with exceptionally high luminosities do not fit with the explosion of a Chandrasekhar mass white dwarf. Moreover, it has been seen that there is a correlation between the surface magnetic field and the mass of white dwarfs. The magnetic white dwarfs seem to be more massive than their nonmagnetic counterparts \cite{Fe15}. Lastly, predictions from the luminosities reveal that the progenitor white dwarfs had  masses significantly higher than the Chandrasekhar limit. It seems that the Chandrasekhar limit may be violated by highly magnetized white dwarfs. To account for these facts, we have calculated theoretically the masses of white dwarfs in presence of such high magnetic fields in the general relativistic formalism.

\noindent
\section{ Completely degenerate ideal Fermi gas and EoS for non-magnetic White Dwarfs }
\label{section2} 

    We consider a relativistic, completely degenerate Fermi gas at zero temperature and neglect any form of interactions between the fermions. By the Pauli exclusion principle, no quantum state can be occupied by more than one fermion with an identical set of quantum numbers. Thus a noninteracting Fermi gas, unlike a Bose gas, is prohibited from condensing into a Bose-Einstein condensate. The total energy of the Fermi gas at absolute zero is larger than the sum of the single-particle ground states because the Pauli principle implies a degeneracy pressure that keeps fermions separated and moving. For this reason, the pressure of a Fermi gas is non-zero even at zero temperature, in contrast to that of a classical ideal gas. This so-called degeneracy pressure stabilizes a white dwarf (a Fermi gas of electrons) against the inward pull of gravity, which would ostensibly collapse the star into a Black Hole. However if a star is sufficiently massive to overcome the degeneracy pressure, it collapse into a singularity due to gravity. While the pressure inside a white dwarf is entirely due to electrons, its mass comes mostly from the atomic nuclei. 

\subsection{ Completely degenerate free Fermi gas }

    The non-interacting assembly of fermions at zero temperature exerts pressure because of kinetic energy from different states filled up to Fermi level. Since pressure is force per unit area which means rate of momentum transfer per unit area, it is given by    
    
\begin{equation}
 P_e= \frac{1}{3}\int p v n_p d^3p = \frac{1}{3}\int \frac{p^2c^2}{\sqrt{(p^2c^2+m_e^2c^4)}} n_p d^3p
\label{seqn3}
\end{equation}
where $m_e$ is the rest mass, $v$ is the velocity of the particles with momentum $\vec p$ and $n_p d^3p$ is the number of particles per unit volume having momenta between $\vec p$ and $\vec p + d\vec p$. The factor $\frac{1}{3}$ accounts for the fact that, on average, only $\frac{1}{3}$rd of total particles $n_p d^3p$ are moving in a particular direction. For fermions having spin $\frac{1}{2}$, degeneracy = 2, $n_p d^3p = \frac{8\pi p^2 dp}{h^3}$ and hence number density $n_e$is given by 

\begin{equation}
 n_e= \int_0^{p_F} n_p d^3p = \frac{8\pi p_F^3}{3h^3} = \frac{x_F^3}{3\pi^2\lambda_e^3}
\label{seqn4}
\end{equation}
where $p_F$ is the Fermi momentum which is maximum momentum possible at zero temperature, $x_F=\frac{p_F}{m_e c}$ is a dimensionless quantity and $\lambda_e=\frac{\hbar}{m_e c}$ is the Compton wavelength. The energy density $\varepsilon_e$ is given by

\begin{equation}
 \varepsilon_e= \int_0^{p_F} E n_p d^3p = \int_0^{p_F} \sqrt{(p^2c^2+m_e^2c^4)} \frac{8\pi p^2 dp}{h^3}
\label{seqn5}
\end{equation}
which along with Eq.(3) turns out upon integration to be

\begin{equation}
 \varepsilon_e= \frac{m_e c^2}{\lambda_e^3} \chi(x_F); ~~~~  P_e= \frac{m_e c^2}{\lambda_e^3} \phi(x_F),
\label{seqn6}
\end{equation}
where

\begin{equation}
 \chi(x)= \frac{1}{8\pi^2}[x\sqrt{1+x^2}(1+2x^2)-\ln(x+\sqrt{1+x^2})]
\label{seqn7}
\end{equation}
and

\begin{equation}
 \phi(x)= \frac{1}{8\pi^2}[x\sqrt{1+x^2}(\frac{2x^2}{3}-1)+\ln(x+\sqrt{1+x^2})].
\label{seqn8}
\end{equation}

\subsection{ EoS for non-magnetic White Dwarfs }

    For the EoS for non-magnetic White Dwarfs, the pressure is provided by the relativistic electrons only and therefore, pressure $P$ is given by 

\begin{equation}
P = P_e= \frac{m_e c^2}{\lambda_e^3} \phi(x_F),
\label{seqn9}
\end{equation}
whereas for energy density $\varepsilon$ both electrons (with its kinetic energy) and atomic nuclei contribute, so that 
    
\begin{equation}
\varepsilon= \varepsilon_e + n_e(m_p+f m_n)c^2 =\frac{m_e c^2}{\lambda_e^3} \chi(x_F) + n_e(m_p+f m_n)c^2 
\label{seqn10}
\end{equation}
where $m_n$ and $m_p$ are the masses of neutron and proton, respectively and $f$ is the number of neutrons per electron. Commonly, electron-degenerate stars consist of helium, carbon, oxygen, etc., for which $f=1$. To be precise, one should in fact also subtract $n_e(1+f)$ times binding energy per nucleon from the second term on the right hand side of the above equation. Obviously, this correction is composition dependent and its contribution being quite small, e.g. in case of helium star it is about 0.7$\%$ to the second term, it is not considered in calculations. Since the kinetic energy of electrons in the above equation contributes negligibly, the mass density for $f=1$ white dwarfs can be expressed in units of 2$\times10^9$ gms/cc by multiplying number density of electrons $n_e$ expressed in units of fm$^{-3}$ by the factor 1.6717305$\times10^6$. 

\noindent
\section{Landau quantization and EoS for magnetized White Dwarfs }
\label{section3}

    Like the former case, here also we consider a completely degenerate relativistic electron gas at zero temperature but embedded in a strong magnetic field. We do not consider any form of interactions with the electrons. Electrons, being charged particles, occupy Landau quantized states in a magnetic field. This changes the EoS, which, in turn, changes the pressure and energy density of the white dwarf. In addition to the matter energy density and pressure, the energy density and pressure due to magnetic field are also taken into account. It is the combined pressure and energy density of matter and magnetic field that determines the mass-radius relation of strongly magnetized white dwarfs. It should be emphasized that protons also, being charged particles, are Landau quantized. But since the proton mass is $\sim 2000$ times the electron mass their cyclotron energy is $\sim 2000$ times smaller than that of the electron for the same magnetic field, and hence we neglect it.

\subsection{ Landau quantization and EoS for free electron gas in magnetic field }

    In order to calculate the thermodynamic quantities like the energy density and pressure of an electron gas in a magnetic field, we need to know the density of states and the dispersion relation. The quantum mechanics of a charged particle in a magnetic field is presented in many texts (e.g. Sokolov and Ternov (1968) \cite{ST68}, Landau and Lifshitz (1977) \cite{LL77}, Canuto and Ventura (1977) \cite{CV77}  M\'esz\'aros (1992) \cite{Me92}). Here we summarize the basics needed for our later discussion. Let us first consider the  motion of a charged particle (charge $q$ and mass $m_e$) in a uniform magnetic field $B$ assumed to be along the z-axis. In classical physics, the particle gyrates in a circular orbit with radius and angular frequency (cyclotron frequency) given by
    
\begin{equation}
 r_c =\frac{m_ecv_{\perp}}{qB};    ~~~~    \omega_c=\frac{qB}{m_ec}
\label{seqn11}
\end{equation}    
where $v_{\perp}$ is the velocity perpendicular to the magnetic field. The hamiltonian of the system is given by  

\begin{equation}
H =\frac{1}{2m_e} \Big(\vec p - \frac{q\vec A}{c}\Big)^2
\label{seqn12}
\end{equation} 
where $\vec B = \nabla \times \vec A$ with $\vec A$ being the electromagnetic vector potential. To have magnetic field in z-direction with magnitude $B$ one must have

\begin{equation}
 \vec A = \bordermatrix{ &\cr 
                        &0 \cr
                        &Bx \cr
                        &0\cr}                      
\label{seqn13}
\end{equation}
and therefore

\begin{equation}
H =\frac{1}{2m_e} [p_x^2 + \Big(p_y - \frac{qBx}{c}\Big)^2 + p_z^2]
\label{seqn14}
\end{equation}
The operator $\hat p_y$ commutes with this hamiltonian since the operator $y$ is absent. Thus operator $\hat p_y$ can be replaced by its eigenvalue $\hbar k_y$.Using cyclotron frequency $\omega_c=\frac{qB}{m_ec}$ one obtains 

\begin{equation}
H =\frac{p_x^2}{2m_e} + \frac{1}{2}m_e\omega_c^2\Big(x - \frac{\hbar k_y}{m_e\omega_c}\Big)^2 + \frac{p_z^2}{2m_e},
\label{seqn15}
\end{equation}
the first two terms of which is exactly the quantum harmonic oscillator with the minimum of the potential shifted in co-ordinate space by $x_0=\frac{\hbar k_y}{m_e\omega_c}$. Noting that translating harmonic oscillator potential does not affect the energies, energy eigenvalues can be given by

\begin{equation}
E_{n ,p_z} =(n+\frac{1}{2})\hbar \omega_c + \frac{p_z^2}{2m_e}, ~~~~n=0, 1, 2 . . . . 
\label{seqn16}
\end{equation}
The energy does not depend on the quantum number $k_y$, so there will be degeneracies. Each set of wave functions with same value of $n$ is called a Landau Level. Each Landau level is degenerate due to the second quantum number $k_y$. If periodic boundary condition is assumed $k_y$ can take values $k_y=\frac{2\pi N}{l_y}$ where $N$ is another integer and $l_x,l_y,l_z$ being the dimensions of the system. The allowed values of $N$ are further restricted by the condition that the centre of the force of the oscillator $x_0$ must physically lie within the system, $0\le x_0 \le l_x$ which implies $0\le N \le \frac{l_x l_y m_e\omega_c }{2\pi\hbar}=\frac{qB l_x l_y}{hc}$. Hence for electrons with spin $s$ and charge $q=-|e|$, the maximum number of particles per Landau level per unit area is $\frac{|e|B(2s+1)}{hc}$. On solving Schr\"odinger's equation for electrons with spin in an external magnetic field in z-direction which is uniform and static, Eq.(16) modifies to 

\begin{equation}
E_{\nu ,p_z}=\nu\hbar \omega_c + \frac{p_z^2}{2m_e}, ~~~~ \nu=n+\frac{1}{2}+s_z.
\label{seqn17}
\end{equation}
Clearly for the lowest Landau level ($\nu=0$) the spin degeneracy $g_\nu=1$ (since only $n=0$, $s_z=-\frac{1}{2}$ is allowed) and for all other higher Landau levels ($\nu\neq0$), $g_\nu=2$ (for $s_z=\pm\frac{1}{2}$).

    For extremely strong magnetic fields such that $\hbar \omega_c \geq m_ec^2$ the motion perpendicular to the magnetic field still remains quantized but becomes relativistic. The solution of the Dirac equation in a constant magnetic field \cite{Lai91} is given by the energy eigenvalues
    
\begin{equation}
E_{\nu ,p_z}=\left[p_z^2c^2+m_e^2c^4\left(1+2\nu B_D\right)\right]^\frac{1}{2}
\label{seqn18}
\end{equation}    
where the dimensionless magnetic field defined as $B_D=B/B_c$ is introduced with $B_c$ given by $\hbar\omega_c=\hbar\frac{|e|B_c}{m_ec}=m_ec^2 \Rightarrow B_c=\frac{m_e^2c^3}{|e|\hbar}=4.414\times 10^{13}$ gauss. Obviously, the density of states in presence of magnetic field gets modified to 

\begin{equation}
\sum\limits_{\nu }\frac{2|e|B}{hc}g_{\nu}\int \frac{dp_z}{h}
\label{seqn19}
\end{equation}
where the sum is on all Landau levels $\nu$. At zero temperature the number density of electrons is given by

\begin{equation}
n_e=\sum\limits_{\nu =0}^{\nu_m} \frac{2|e|B}{h^2c} g_{\nu} \int_0^{p_F(\nu )} dp_z = \sum\limits_{\nu =0}^{\nu_m}\frac{2|e|B}{h^2c}g_{\nu}p_F(\nu)
\label{seqn20}
\end{equation}
where $p_F(\nu )$ is the Fermi momentum in the $\nu $th Landau level and $\nu_m$ is the upper limit of the Landau level summation. The Fermi energy $E_F$ of the $\nu $th Landau level is given by

\begin{equation}
E_F^2=p_F^2(\nu)c^2+m_e^2c^4\left(1+2\nu B_D\right)
\label{seqn21}
\end{equation}
and $\nu_m$ can be found from the condition $[p_F(\nu)]^2 \geq 0$ or

\begin{equation}
\nu \leq \frac{\epsilon_F^2-1}{2B_D} ~~\Rightarrow ~~ \nu_m =\frac{\epsilon_{F max}^2-1}{2B_D},
\label{seqn22}
\end{equation}     
where $\epsilon_F=\frac{E_F}{m_ec^2}$ is the dimensionless Fermi energy and $\epsilon_{F max}=\frac{E_{F max}}{m_ec^2}$ the dimensionless maximum Fermi energy of a system for a given $B_D$ and $\nu_m$. Obviously, very small $B_D$ corresponds to large number of Landau levels leading to the familiar non-magnetic EoS. $\nu_m$ is taken to be the nearest lowest integer. Like the former case, if we define a dimensionless Fermi momentum $x_F(\nu)=\frac{p_F(\nu )}{m_ec}$ then Eqns.(20) and (21) may be written as

\begin{equation}
n_e=\frac{2B_D}{(2\pi )^2\lambda_e^3}\sum\limits_{\nu =0}^{\nu_m} g_{\nu}x_F(\nu )
\label{seqn23}
\end{equation}
and

\begin{equation}
\epsilon_F=\left[x^2_F(\nu )+1+2\nu B_D\right]^\frac{1}{2}
\label{seqn24}
\end{equation}
or

\begin{equation}
x_F(\nu )=\left[\epsilon_F^2-(1+2\nu B_D)\right]^\frac{1}{2}.
\label{seqn25}
\end{equation}   
The electron energy density is given by

\begin{eqnarray}
\varepsilon_e&&=\frac{2B_D}{(2\pi )^2\lambda_e^3}\sum\limits_{\nu =0}^{\nu_m} g_{\nu }\int_0^{x_F(\nu )}E_{\nu ,p_z}d\left(\frac{p_z}{m_ec}\right) \nonumber\\
&&=\frac{2B_D}{(2\pi )^2\lambda_e^3}m_ec^2\sum\limits_{\nu =0}^{\nu_m} g_{\nu }(1+2\nu B_D)\psi \left(\frac{x_F(\nu )}{(1+2\nu B_D)^{1/2}}\right), \nonumber\\
\label{seqn26}
\end{eqnarray}
where

\begin{equation}
\psi (z)=\int_0^z(1+y^2)^{1/2}dy=\frac{1}{2}[z\sqrt{1+z^2}+\ln(z+\sqrt{1+z^2})]
\label{seqn27}
\end{equation}
The pressure of the electron gas is given by

\begin{eqnarray}
P_e&&=n_e^2\frac{d}{dn_e}\left(\frac{\varepsilon_e}{n_e}\right)= n_eE_F -\varepsilon_e \nonumber\\
&&=\frac{2B_D}{(2\pi )^2\lambda_e^3}m_ec^2\sum\limits_{\nu =0}^{\nu_m} g_{\nu }(1+2\nu B_D)\eta \left(\frac{x_F(\nu )}{(1+2\nu B_D)^{1/2}}\right), \nonumber\\
\label{seqn28}
\end{eqnarray}
where

\begin{equation}
\eta (z)=z\sqrt{1+z^2}-\psi (z) =\frac{1}{2}[z\sqrt{1+z^2}-\ln(z+\sqrt{1+z^2})].
\label{seqn29}
\end{equation}

\subsection{ Magnetized White Dwarfs }
 
    In the present case of magnetic White Dwarfs, the explicit contributions from the energy density $\varepsilon_B=\frac{B^2}{8\pi}$ and pressure $P_B=\frac{1}{3}\varepsilon_B$ arising due to magnetic field need to be added to the matter energy density and pressure as 
    
\begin{eqnarray}
P=&&P_e+P_B  \nonumber\\
=&&\frac{2B_D}{(2\pi )^2\lambda_e^3}m_ec^2\sum\limits_{\nu =0}^{\nu_m} g_{\nu }(1+2\nu B_D)\eta \left(\frac{x_F(\nu )}{(1+2\nu B_D)^{1/2}}\right)  \nonumber\\
&&+\frac{B^2}{24\pi},
\label{seqn30}
\end{eqnarray}
and
  
\begin{eqnarray}
\varepsilon=&& \varepsilon_e + n_e(m_p+f m_n)c^2+\varepsilon_B   \nonumber\\
=&&\frac{2B_D}{(2\pi )^2\lambda_e^3}m_ec^2\sum\limits_{\nu =0}^{\nu_m} g_{\nu }(1+2\nu B_D)\psi \left(\frac{x_F(\nu )}{(1+2\nu B_D)^{1/2}}\right)  \nonumber\\
&&+ n_e(m_p+f m_n)c^2  +\frac{B^2}{8\pi}.
\label{seqn31}
\end{eqnarray}

\noindent
\section{Tolman-Oppenheimer-Volkoff Equation and theoretical calculations of mass-radius relation for White Dwarfs}
\label{section4}
    
    If rapidly rotating compact stars were non-axisymmetric, they would emit gravitational waves in a very short time scale and settle down to axisymmetric configurations. Therefore, we need to solve for rotating and
axisymmetric configurations in the framework of general relativity. For the matter and the spacetime the following assumptions are made. The matter distribution and the spacetime are axisymmetric, the matter and the spacetime are in a stationary state, the matter has no meridional motions, the only motion of the matter is a circular one that is represented by the angular velocity, the angular velocity is constant as seen by a distant observer at rest and the matter can be described as a perfect fluid. To study the rotating stars the following metric is used

\vspace{-0.5cm}
\begin{eqnarray}
ds^2 = -e^{(\gamma+\rho)} dt^2 + e^{2\alpha} (dr^2+r^2d\theta^2) \nonumber\\
       + e^{(\gamma-\rho)} r^2 \sin^2\theta (d\phi-\omega dt)^2
\label{seqn32}
\end{eqnarray}
\noindent
where the gravitational potentials $\gamma$, $\rho$, $\alpha$ and $\omega$ are functions of polar coordinates $r$ and $\theta$ only. The Einstein's field equations for the three potentials $\gamma$, $\rho$ and $\alpha$ can be solved using the Green's-function technique \cite{Ko89} and the fourth potential $\omega$ can be determined from other potentials. All the physical quantities may then be determined from these potentials. Rotational frequency of stars is limited by Kepler's frequency which is the mass shedding limit. For very compact stars such as neutron stars the Kepler's frequency is very high and can go up to millisecond order \cite{Ch10,Mi12} whereas white dwarfs being about thousand times bigger in size and much less dense, Kepler's frequency is very small and one may safely use the zero frequency limit \cite{Ba14} to the Einstein's field equations. Obviously, at the zero frequency limit corresponding to the static solutions of the Einstein's field equations for spheres of fluid, the present formalism yields the results for the solution of the Tolman-Oppenheimer-Volkoff (TOV) equation \cite{TOV39a,TOV39b} given by
 
\vspace{-0.5cm}
\begin{eqnarray}
\frac{dP(r)}{dr} = -\frac{G}{c^4}\frac{[\varepsilon(r)+P(r)][m(r)c^2+4\pi r^3P(r)]}{r^2[1-\frac{2Gm(r)}{rc^2}]} \\ 
{\rm where} ~~\varepsilon(r)=\rho(r)c^2~~{\rm and} ~~m(r)c^2=\int_0^r \varepsilon(r') d^3r' \nonumber
\label{seqn33}
\end{eqnarray}
\noindent   
which can be easily solved numerically using Runge-Kutta method for masses and radii. The quantities $\varepsilon(r)$ and $P(r)$ are the energy density and pressure at a radial distance $r$ from the centre, and are given by the equation of state. The mass of the star contained within a distance $r$ is given by $m(r)$. The size of the star is determined by the boundary condition $P(r)=0$ and the total mass $M$ of the star integrated up to the surface $R$ is given by $M=m(R)$ \cite{Um97}. The single integration constant needed to solve the TOV equation is $P_c$, the pressure at the center of the star calculated at a given central density $\rho_c$.
    
\noindent
\section{ Results and discussion }
\label{section5}

    Recently, there are some important calculations for masses and radii of magnetized white dwarfs using non-relativistic Lane-Emden equation assuming a constant magnetic field throughout which provided masses up to 2.3-2.6 $M_\odot$ \cite{Das12}, a mass significantly greater than the Chandrasekhar limit. However, because of the structure of the Lane-Emden equation, pressure arising due to constant magnetic field do not contribute while for the general relativistic TOV equation case is not the same. Moreover, the EoS needed to be fitted to a polytropic form. In order to derive a mass limit for magnetized white dwarfs (similar to the mass limit of $\sim$1.4 $M_\odot$ obtained by Chandrasekhar \cite{Ch35} for non-magnetic white dwarfs), the same authors, under certain approximations, have been able to reduce the EoS to a polytropic form with index $1+1/n=2$ for which analytic solution of Lane-Emden equation exists ($\theta(\xi)=sin\xi/\xi$ where $\rho=\rho_c\theta^n$ with $\rho$ and $\rho_c$ being density and central density, respectively) and avoiding the energy density $\varepsilon_B=\frac{B^2}{8\pi}$ and pressure $P_B=\frac{1}{3}\varepsilon_B$ arising due to magnetic field by assuming it to be constant throughout, they were able to set a mass limit of 2.58 $M_\odot$ \cite{Das13,Do14}. In the present work, we have calculated masses and radii of white dwarfs by solving the general relativistic TOV equation both for non-magnetic and magnetized white dwarfs using the exact EoS without resorting to fit it to a polytropic form. 

\subsection{Chandrasekhar limit for White Dwarfs}

    We verify Chandrasekhar limit \cite{Ch35} for masses of white dwarfs by actually solving TOV equation for non-magnetic white dwarfs. The masses and radii of such white dwarfs are listed in Table-I. It is interesting to note that considering a very high central density of 3.343$\times10^{10}$ gms/cc for $f=1$ white dwarfs, one can asymptotically reach the Chandrasekhar mass limit. It is important to mention that beyond this density at $\sim$ 4.3$\times10^{11}$ gms/cc, the 'neutron drip' point \cite{Ch15}, the nuclei become so neutron rich that with increasing density the continuum neutron states begin to be filled, and the lattice of neutron-rich nuclei becomes permeated by a sea of neutrons. In Table-I, masses and radii of non-magnetic white dwarfs as a function of central density are provided. In Fig.-1 plot for masses of non-magnetic white dwarfs is shown as a function of central density whereas in Fig.-2 mass-radius relationship of non-magnetic white dwarfs is provided. These results for non-magnetic white dwarfs do conform to the traditional Chandrasekhar mass-limit.
    
\begin{table}[h!]
\centering
\caption{Variations of masses and radii of non-magnetic white dwarfs with central number density of electrons which can be expressed in units of 2$\times10^9$ gms/cc for mass density by multiplying with 1.6717305$\times10^6$.}
\begin{tabular}{||c|c|c||}
\hline 
\hline
$~~~~n_e$ (r=0)$~~~~$&$~~~~$Radius$~~~~$ &$~~~~$Mass$~~~~$ \\ \hline
fm$^{-3}$&Kms & $M_\odot$ \\ \hline
\hline
1.0$\times$10$^{-5}$&917.87&1.3904 \\
5.0$\times$10$^{-6}$&1126.83&1.3905 \\
4.0$\times$10$^{-6}$&1202.53&1.3896 \\
3.8$\times$10$^{-6}$&1220.55&1.3893 \\
3.6$\times$10$^{-6}$&1239.80&1.3890 \\
3.4$\times$10$^{-6}$&1260.43&1.3887 \\
3.2$\times$10$^{-6}$&1282.65&1.3883 \\
3.0$\times$10$^{-6}$&1306.67&1.3878 \\
2.8$\times$10$^{-6}$&1332.78&1.3873 \\
2.6$\times$10$^{-6}$&1361.33&1.3866 \\
2.4$\times$10$^{-6}$&1392.75&1.3859 \\
2.2$\times$10$^{-6}$&1427.62&1.3850 \\
2.0$\times$10$^{-6}$&1466.67&1.3839 \\
1.8$\times$10$^{-6}$&1510.90&1.3825 \\
1.6$\times$10$^{-6}$&1561.69&1.3809 \\
1.4$\times$10$^{-6}$&1621.01&1.3788 \\
1.2$\times$10$^{-6}$&1691.86&1.3761 \\
1.0$\times$10$^{-6}$&1779.00&1.3724 \\
8.0$\times$10$^{-7}$&1890.72&1.3673 \\
6.0$\times$10$^{-7}$&2043.29&1.3594 \\
4.0$\times$10$^{-7}$&2275.36&1.3457 \\
2.0$\times$10$^{-7}$&2721.16&1.3138 \\
1.0$\times$10$^{-7}$&3233.63&1.2692 \\
1.0$\times$10$^{-8}$&5482.58&1.0051 \\
1.0$\times$10$^{-9}$&8721.75&0.5949 \\ \hline
\hline
\end{tabular}
\label{table1} 
\end{table}
\noindent 

\begin{figure}[t]
\vspace{0.0cm}
\eject\centerline{\epsfig{file=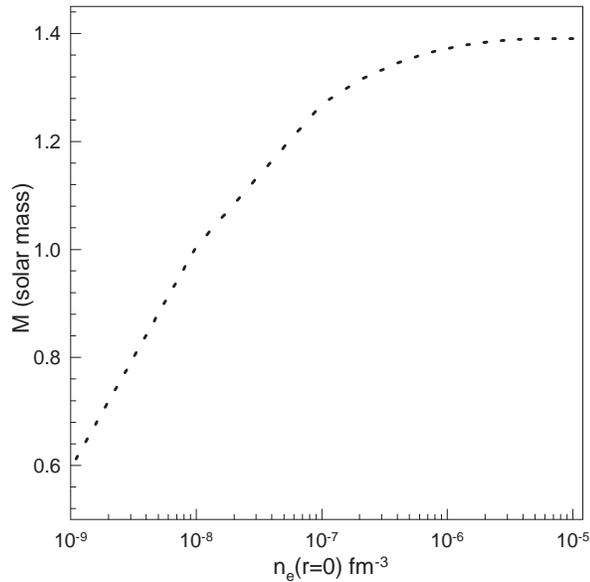,height=7.7cm,width=7.7cm}}
\caption
{Plot for masses of non-magnetic white dwarfs as a function of central density.}
\label{fig1}
\vspace{0.0cm}
\end{figure}
\noindent 

\subsection{Super-Chandrasekhar White Dwarfs}

    As mentioned in the beginning of this section that unlike non-relativistic Lane-Emden equation, pressure arising due to constant magnetic field does contribute to the general relativistic TOV equation. Presence of high constant magnetic field do not provide valid solutions to the TOV equations. Hence, we present stable solutions of magnetostatic equilibrium models for super-Chandrasekhar white dwarfs with varying magnetic field profiles which is maximum at the centre and goes to 10$^{9}$ gauss at the surface of the star. This has been obtained by self-consistently including the effects of the magnetic pressure gradient and total magnetic density in a general relativistic framework. Nevertheless, we have also performed calculations corresponding to very high (single Landau level) and high (multi Landau levels) magnetic field which is constant throughout the star in order to compare with the results from solutions of Lane-Emden equation described above, but for these cases we have to ignore the explicit contributions from energy density $\varepsilon_B$ and pressure $P_B$ arising due to magnetic field. Results of such calculations are provided in Tables-II $\&$ III for magnetized white dwarfs with single and multiple Landau levels, respectively.

\begin{figure}[t]
\vspace{0.0cm}
\eject\centerline{\epsfig{file=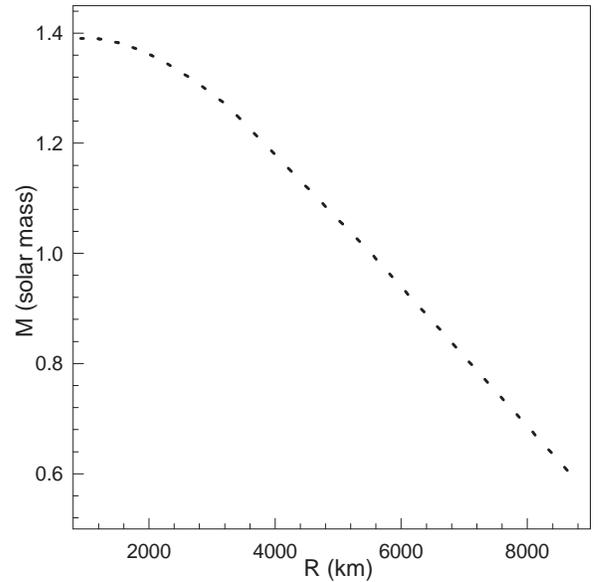,height=7.7cm,width=7.7cm}}
\caption
{Plot for mass-radius relationship of non-magnetic white dwarfs.}
\label{fig2}
\vspace{0.0cm}
\end{figure}
\noindent 

\begin{table}[htbp]
\centering
\caption{Variations of masses and radii of uniformly magnetized white dwarfs with central number density of electrons which can be expressed in units of 2$\times10^9$ gms/cc for mass density by multiplying with 1.6717305$\times10^6$. The minimum magnetic field $B_{dmin}$ corresponding to the central density required to make single Landau level throughout is listed in units of $B_c$.}
\begin{tabular}{||c|c|c|c||}
\hline 
\hline
$~~~~n_e$ (r=0)$~~~~$&$~~~~$Radius$~~~~$ &$~~~~$Mass$~~~~$&$~~~~$$B_{dmin}$$~~~~$ \\ \hline
fm$^{-3}$&Kms & $M_\odot$ &in units of $B_c$ \\ \hline
\hline
5.0$\times$10$^{-6}$&592.28&2.4521&253 \\
4.0$\times$10$^{-6}$&636.54&2.4508&218 \\
3.0$\times$10$^{-6}$&698.11&2.4461&180 \\
2.0$\times$10$^{-6}$&792.71&2.4204&138 \\
1.0$\times$10$^{-6}$&989.84&2.4149&86.5 \\ \hline
\hline
\end{tabular}
\label{table2} 
\end{table}
\noindent     
   
\begin{figure}[t]
\vspace{0.0cm}
\eject\centerline{\epsfig{file=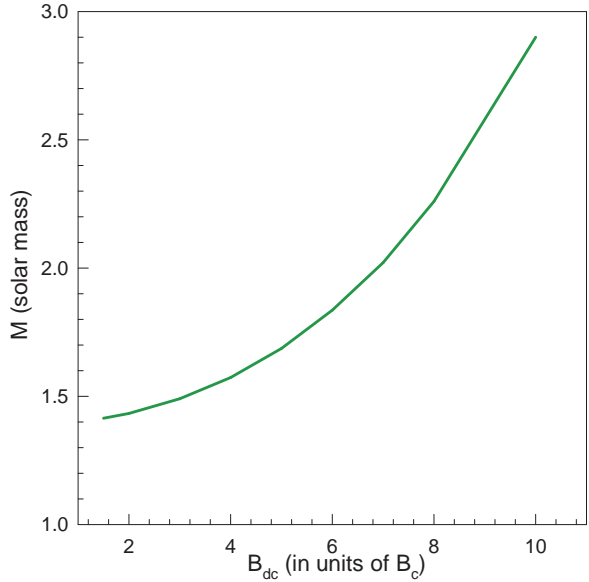,height=7.7cm,width=7.7cm}}
\caption
{Plot for masses of magnetized white dwarfs as a function of central magnetic field.}
\label{fig3}
\vspace{0.0cm}
\end{figure}
\noindent     
   
\begin{table}[htbp]
\centering
\caption{Variations of masses and radii of uniformly magnetized white dwarfs with central number density of electrons which can be expressed in units of 2$\times10^9$ gms/cc for mass density by multiplying with 1.6717305$\times10^6$. The magnetic field $B_d$ ($<$ $B_{dmin}$ for the central density) is also listed which generates multiple Landau levels.}
\begin{tabular}{||c|c|c|c||}
\hline 
\hline
$~~~~n_e$ (r=0)$~~~~$&$~~~~$Radius$~~~~$ &$~~~~$Mass$~~~~$&$~~~~$$B_d$$~~~~$ \\ \hline
fm$^{-3}$&Kms & $M_\odot$ &in units of $B_c$\\ \hline
\hline
4.6736$\times$10$^{-6}$&1149.77&1.3925&1.5 \\
3.5147$\times$10$^{-6}$&663.58&2.4491&200 \\ \hline
\hline
\end{tabular}
\label{table3} 
\end{table}
\noindent     
 
    Now we perform the actual calculations with varying magnetic field including the effects of energy density and pressure arising due to magnetic field in a general relativistic framework. The variation of magnetic field \cite{Ba97} inside white dwarf is taken to be of the form   
 
\begin{equation}
 B_d = B_s + B_0[1-\exp\{-\beta(n_e/n_0)^\gamma\}]
 \label{seqn34}
\end{equation} 
where $B_d$ (in units of $B_c$) is the magnetic field at electronic number density $n_e$, $B_s$ (in units of $B_c$) is the surface magnetic field and $n_0$ is taken as $n_e$(r=0)/10 and $\beta$, $\gamma$ are constants. Once central magnetic field is fixed, $B_0$ can be determined from above equation. We choose constants $\beta=0.8$ and $\gamma=0.9$, rather arbitrarily by using unequal non-unity values, which provides stable solutions of magnetostatic equilibrium models for super-Chandrasekhar white dwarfs. Nevertheless, the magnetic field is not taken completely in ad hoc manner, because central and surface magnetic fields once fixed the variations of its profile do not alter the gross results. Moreover, we have kept maximum central magnetic field strength at 10$B_c$ which is $4.414 \times 10^{14}$ gauss, near to the lower of the maximum limit  suggested by N. Chamel et al. \cite{Ch13} and surface magnetic field $\sim10^{9}$ gauss estimated by observations. In Table-IV the results of these realistic calculations are listed. In Figs.-3,4 plots for masses and radii of magnetized white dwarfs  are shown as functions of central magnetic field. Present calculations estimate that the maximum stable mass of magnetized white dwarfs could be $\sim$3 $M_\odot$. These results are quite useful in explaining the peculiar, overluminous type Ia supernovae that do not conform to the traditional Chandrasekhar mass-limit.   
 
\begin{figure}[t]
\vspace{0.0cm}
\eject\centerline{\epsfig{file=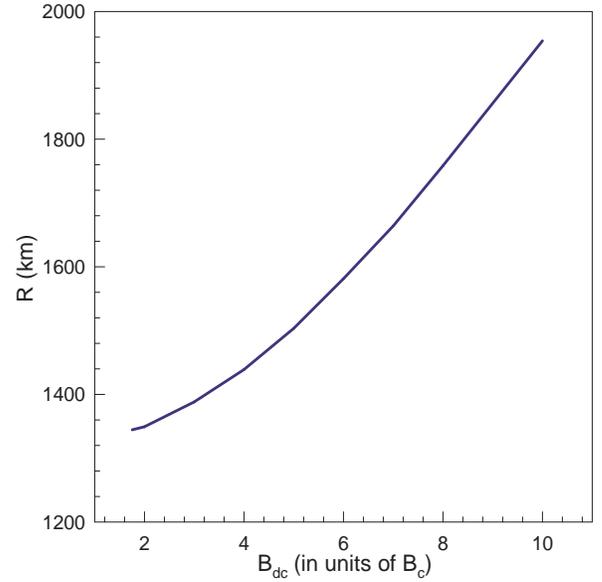,height=7.7cm,width=7.7cm}}
\caption
{Plot for radii of magnetized white dwarfs as a function of central magnetic field.}
\label{fig4}
\vspace{0.0cm}
\end{figure}
\noindent 
   
\noindent
\section{ Summary and conclusion }
\label{section6}
    
    In summary, we have considered a relativistic, degenerate electron gas at zero temperature under the influence of a density dependent magnetic field. Since the electrons are Landau quantized, the density of states gets modified due to the presence of the magnetic field. This, in turn, modifies the EoS of the white dwarf matter. The presence of magnetic field also gives rise to magnetic energy density and pressure which is added to those due to degenerate matter. We find that the masses of such white dwarfs increase with the magnitude of the central magnetic field. Hence we obtain a conclusive result that it is possible to have 
electron-degenerate magnetized white dwarfs, with masses significantly greater than the Chandrasekhar limit in the range of ~$\sim$3 $M_\odot$, provided it has an appropriate magnetic field profile with high magnitude at the centre as well as high central density.                    
              
\begin{table}[htbp]
\centering
\caption{Variations of masses and radii of magnetized white dwarfs with central number density of electrons which can be expressed in units of 2$\times10^9$ gms/cc for mass density by multiplying with 1.6717305$\times10^6$. The maximum magnetic field $B_{dc}$ at the centre is listed in units of $B_c$ whereas the surface magnetic field $B_s$ is taken to be 10$^{9}$ gauss.}
\begin{tabular}{||c|c|c|c||}
\hline 
\hline
$~~~~n_e$ (r=0)$~~~~$&$~~~~$Radius$~~~~$ &$~~~~$Mass$~~~~$&$~~~~$$B_{dc}$$~~~~$ \\ \hline
fm$^{-3}$&Kms & $M_\odot$ &in units of $B_c$ \\ \hline
\hline
4.674017$\times$10$^{-6}$&1285.91&1.4146&1.5 \\
4.673846$\times$10$^{-6}$&1344.46&1.4236&1.75 \\
4.674209$\times$10$^{-6}$&1349.45&1.4339&2.0 \\
4.675374$\times$10$^{-6}$&1388.04&1.4906&3.0 \\
4.672188$\times$10$^{-6}$&1438.94&1.5731&4.0 \\
4.670830$\times$10$^{-6}$&1503.64&1.6863&5.0 \\
4.678118$\times$10$^{-6}$&1581.27&1.8353&6.0 \\
4.677677$\times$10$^{-6}$&1663.86&2.0217&7.0 \\
4.665741$\times$10$^{-6}$&1758.40&2.2601&8.0 \\
4.661657$\times$10$^{-6}$&1954.44&2.8997&10. \\ \hline
\hline
\end{tabular}
\label{table4} 
\end{table}
\noindent 
                    
    To date there are about $\sim$250 magnetized white dwarfs with well determined fields \cite{Fe15}
and over $\sim$600 if objects with no or uncertain field determination \cite{Ke13,Ke15} are also included. Surveys such as the SDSS, HQS and the Cape Survey have discovered these magnetized white dwarfs. The complete samples show that the field distribution of magnetized white dwarfs is in the range 10$^3$-10$^9$ gauss which basically provides the surface magnetic fields. However, the central magnetic field strength, which is presumably unobserved by the above observations, could be several orders of magnitude higher than the surface field. In fact, it is the central magnetic field which is crucial for super-Chandrasekhar magnetized white dwarfs. However, the softening of the EoS accompanying the onset of electron captures and pycnonuclear reactions in the core of these stars can lead to local instabilities which set an upper limit to the magnetic field strength at the center of the star, ranging from 10$^{14}$-10$^{16}$ gauss depending on the core \cite{Ch13} composition.


\begin{thebibliography}{99}

\bibitem{Vi15} Ivan Marti-Vidal, Sebastien Muller, Wouter Vlemmings, Cathy Horellou, Susanne Aalto, Science {\bf 348}, 6232 (2015).

\bibitem{Ho06} D. Andrew Howell et al., Nature {\bf 443}, 308 (2006).

\bibitem{Sc10} R. A. Scalzo et al., Astrophys. J. {\bf 713}, 1073 (2010).

\bibitem{Hi07} M. Hicken et al., Astrophys. J. {\bf 669}, L17 (2007).

\bibitem{Ya09} M. Yamanaka et al., Astrophys. J. {\bf 707}, L118 (2009).

\bibitem{Si11} Jeffrey M. Silverman et al., Mon. Not. R. Astron. Soc., {\bf 410}, 1 (2011). 

\bibitem{Fe15} Lilia Ferrario, Domitilla deMartino, Boris T. Gansicke, arXiv: 1504.08072v1 (2015). 

\bibitem{ST68} A. A. Sokolov and I. M. Ternov, {\it Synchrotron Radiation} (Pergamon, Oxford) (1968).

\bibitem{LL77} L. D. Landau and E. M. Lifshitz, {\it Quantum Mechanics} (3rd. ed.; Oxford: Pergamon) (1977).

\bibitem{CV77} V. Canuto and J. Ventura, Fundam. Cosm. Phys. {\bf 2}, 203 (1977).

\bibitem{Me92} P. M\'esz\'aros, {\it High Energy Radiation from Magnetized Neutron Stars} (University of Chicago, Chicago) (1992).

\bibitem{Lai91} Dong Lai and Stuart L. Shapiro, Astrophys. J. {\bf 383}, 745 (1991).

\bibitem{Ko89} H. Komatsu, Y. Eriguchi, I. Hachisu, Mon. Not. R. Astron. Soc. {\bf 237}, 355 (1989).

\bibitem{Ch10} P. R. Chowdhury, A. Bhattacharyya, D. N. Basu, Phys. Rev. {\bf C 81}, 062801(R) (2010).

\bibitem{Mi12} Abhishek Mishra, P. R. Chowdhury and  D. N. Basu, Astropart. Phys. {\bf 36}, 42 (2012).

\bibitem{Ba14} D. N. Basu, Partha Roy Chowdhury and Abhishek Mishra, Eur. Phys. J. Plus {\bf 129}, 62 (2014). 

\bibitem{TOV39a} R. C. Tolman, Phys. Rev. {\bf 55}, 364 (1939).

\bibitem{TOV39b} J. R. Oppenheimer and G. M. Volkoff Phys. Rev. {\bf 55}, 374 (1939).

\bibitem{Um97} V. S. Uma Maheswari, D. N. Basu, J. N. De and S. K. Samaddar, Nucl. Phys. {\bf A 615}, 516 (1997).

\bibitem{Das12} U. Das and B. Mukhopadhyay, Phys. Rev. {\bf D 86}, 042001 (2012).

\bibitem{Ch35} S. Chandrasekhar, Mon. Not. R. Astron. Soc. {\bf 95}, 207 (1935).

\bibitem{Das13} U. Das and B. Mukhopadhyay, Phys. Rev. Lett. {\bf 110}, 071102 (2013).

\bibitem{Do14} J. M. Dong, W. Zuo, P. Yin and J. Z. Gu, Phys. Rev. Lett. {\bf 112}, 039001 (2014).

\bibitem{Ch15} N. Chamel, Zh. K. Stoyanov, L. M. Mihailov, Y. D. Mutafchieva, R. L. Pavlov and Ch. J. Velchev, Phys. Rev. {\bf C 91}, 065801 (2015).

\bibitem{Ba97} D. Bandyopadhyay, S. Chakrabarty and S. Pal, Phys. Rev. Lett. {\bf 79}, 2176 (1997).

\bibitem{Ch13} N. Chamel, A. F. Fantina and P. J. Davis, Phys. Rev. {\bf D 88}, 081301(R) (2013).

\bibitem{Ke13} S. O. Kepler, I. Pelisoli, S. Jordan, S. J. Kleinman, D. Koester, B. K\"ulebi, V. Pecanha,
B. G. Castanheira, A. Nitta, J. E. S. Costa, D. E. Winget, A Kanaan,  and L. Fraga, Mon. Not. R. Astron. Soc. {\bf 429}, 2934 (2013).

\bibitem{Ke15} S. O. Kepler, I. Pelisoli, D. Koester, G. Ourique, S. J. Kleinman, A. D. Romero, A. Nitta, D. J. Eisenstein, J. E. S. Costa, B. K\"ulebi, S. Jordan, P. Dufour, P. Giommi and A. Rebassa-Mansergas, Mon. Not. R. Astron. Soc. {\bf 446}, 4078 (2015).

\end{thebibliography}
\end{document}